\documentclass{aa}

\usepackage{lscape}
\usepackage{natbib}
\bibpunct{(}{)}{;}{a}{}{,} 
\usepackage{graphicx}
\usepackage{txfonts}
\usepackage{color,soul}
\begin{document}

   \title{A missing link in the nitrogen-rich organic chain on Titan}


   \author{N. Carrasco
          \inst{1}
          \and
          J. Bourgalais\inst{1}\fnmsep\thanks{Now at Université de Lorraine, CNRS, LRGP, F-54000 Nancy, France}
          \and 
          L. Vettier\inst{1}
          \and
           P. Pernot\inst{2}
          \and
          E. Giner\inst{3}
          \and 
          R. Spezia\inst{3}
          }

   \institute{Université Paris-Saclay, UVSQ, CNRS, LATMOS, 78280, Guyancourt, France\\
              \email{nathalie.carrasco@latmos.ipsl.fr}
         \and
            Université Paris-Saclay, CNRS, Institut de Chimie Physique, 91405, Orsay, France\\
         \and
             Sorbonne Université, Laboratoire de Chimie Théorique, UMR 7616 CNRS,  75005, Paris, France\\
             \email{riccardo.spezia@sorbonne-universite.fr}
             }
             


   \date{Received XX, 20XX; accepted XX XX, 20XX}

 
  \abstract
   {The chemical building blocks of life contain a large proportion of nitrogen, an essential element. Titan, the largest moon of Saturn, with its dense atmosphere of molecular nitrogen and methane, offers an exceptional opportunity to explore how this element is incorporated into carbon chains through atmospheric chemistry in our Solar System. A brownish dense haze is consistently produced in the atmosphere and accumulates on the surface on the moon. This solid material is nitrogen-rich and may contain prebiotic molecules carrying nitrogen.}
   {To date, our knowledge of the processes leading to the incorporation of nitrogen into organic chains has been rather limited. In the present work, we investigate the formation of nitrogen-bearing ions in an experiment simulating Titan's upper atmosphere, with strong implications for the incorporation of nitrogen into organic matter on Titan.}
   {By combining experiments and theoretical calculations, we show that the abundant N$_2^+$ ion, produced at high altitude by extreme-ultraviolet solar radiation, is able to form nitrogen-rich organic species.}
   {An unexpected and important formation of CH$_3$N$_2^+$ and CH$_2$N$_2^+$ diazo-ions is experimentally observed when exposing a gas mixture composed of molecular nitrogen and methane to extreme-ultraviolet radiation. Our theoretical calculations show that these diazo-ions are mainly produced by the reaction of N$_2^+$ with CH$_3$ radicals. These small nitrogen-rich diazo-ions, with a N/C ratio of two, appear to be a missing link that could explain the high nitrogen content in Titan’s organic matter. More generally, this work highlights the importance of reactions between ions and radicals, which have rarely been studied thus far, opening up new perspectives in astrochemistry.}
   {}

   \maketitle
%

\section{Introduction}

  The NASA-ESA Cassini-Huygens space mission visited Titan, the largest moon of Saturn, for thirteen years between 2004 and 2017. The mission revealed that Titan is a world of organic matter. An organic haze is produced by photochemistry in the atmosphere and surrounds the satellite at all times \citep{RN1}. At the poles, extended organic lakes and seas have been found \citep{RN2,RN3}, while organic sand dunes accumulate at the equator \citep{RN4}. Organic chemistry is basically carbon chemistry and is supported on Titan by the presence of large amounts of methane (between 0.5 and 3.4 \% in the stratosphere from the infrared interferometer spectrometer \citep{RN34}), which partially dissociates at high altitudes through the effect of solar ultraviolet radiation. The radicals and ions thus produced recombine and form heavier hydrocarbons such as ethane, acetylene, and benzene \citep{RN5,RN6,RN7}, as well as poly-aromatic-hydrocarbons \citep{RN8,RN9}, which are large organic structures that are also detected in circumstellar and interstellar environments \citep{RN10}. Carbon chemistry is a prerequisite for the advent of life on the early Earth, but life also requires the contribution of heteroatoms, such as oxygen, nitrogen, sulfur, and phosphorus. Among them, nitrogen makes up part of the structure of proteins and DNA bases. Adenine, C$_5$H$_5$N$_5$, one of the four bases of DNA, has as many nitrogen atoms as carbon, showing an incredibly high incorporation of nitrogen into its structure. The ways of coupling nitrogen into organic chemistry to support the advent of life has not gained a broad consensus. It potentially involves a high concentration of hydrogen cyanide on the early Earth \citep{RN11}. However, Titan’s atmosphere, mainly composed of molecular nitrogen N$_2$ and methane CH$_4$, provides an ideal atmosphere to explore the coupling between the chemistry of carbon and nitrogen. \par 
  The Huygens probe aboard Cassini analyzed the chemical composition of Titan haze grains after the pyrolysis of their refractory nucleus. Intense features of hydrogen cyanide and ammonia were detected in addition to acetylene, suggesting the efficiency in incorporating nitrogen atoms into the complex macromolecules composing the haze grains \citep{RN12}. The high level of incorporation for nitrogen into these organic structures is puzzling and cannot be explained by current knowledge of gas-phase chemistry. Neutral photochemistry modeling sustains the formation of numerous mononitriles (HCN and the isonitrile HNC, HC$_3$N, HC$_5$N, CH$_3$CN, C$_2$H$_5$CN, C$_2$H$_3$CN, C$_3$H$_7$CN, CH$_3$C$_3$N, C$_4$H$_3$CN, and C$_4$H$_5$CN), as well as the two dinitriles (C$_2$N$_2$ and C$_4$N$_2$), the simplest imine CH$_2$NH, and amine CH$_3$NH$_2$ \citep{RN15}. The highest N/C ratio achieved by models is 1 and is limited to molecule sizes not exceeding C$_2$N$_2$. Carbon chain growth processes produce larger molecules, while nitrogen is incorporated by the single addition of a nitrogen atom or a CN radical. A similar conclusion has  been drawn for the chemistry of positive and negative ions: the reaction pathways to explain the incorporation of nitrogen are very limited and all expected ions involve at most one nitrogen atom \citep{RN16,RN17}. Most known organic growth processes for ion chemistry are currently based on the addition of C$_2$H$_2$ to small ions \citep{RN18}. Chemical reactions are missing in the models to explain the large nitrogen content in Titan aerosols and their laboratory analogues. Based on the knowledge that the first nanoparticles composing Titan's haze are produced in the thermosphere, where molecular nitrogen, the major compound, is partially ionized to N$_2^+$, the aim of this work is to study whether the reactivity of N$_2^+$ could assist the incorporation of nitrogen into organic molecules via efficient ion-neutral reactions and, more specifically, via ion-radical reactions. \par 
There are indeed few measurements of ion-neutral reactions that include polyatomic free radicals, especially hydrocarbon-based species \citep{RN19,RN20}. The explanation lies in the difficulty of experimental studies. Experimental studies of these reactions remain difficult, mainly due to the difficulty of generating the radicals cleanly and intensely. Radicals have to be produced in situ from the destruction of a precursor, which often leads to  low quantities. The precursor can react faster with the ion than the radical itself. This also leads to problems of radical recombination and control of the internal states of the reagents that influence their reactivity. \par
 Quantum chemistry calculations are a very useful tool to propose possible reactions and thus to explain the formation of diazo compounds. Here, they were conducted not only on the suggested reaction of N$_2^+$ ion with the CH$_3$ radical, but also on the possible reaction of neutral N$_2$ with CH$_3^+$ and CH$_4^+$ ions, to identify which is most likely both thermodynamically and kinetically.\par
In the atmosphere of Titan, the interactions with solar photons are the most efficient processes to activate the chemistry based on N$_2$ and CH$_4$ \citep{lavvas2011energy}. Exposure to solar flux triggers efficient photochemistry of the most abundant species (N$_2$ and CH$_4$), leading to the formation of radical and ionic species through photoionization and photodissociation.

Kinetics of ion-neutral reactions is few sensitive to temperature, as most reactions are already fast and proceed at rate values close to the collision rate (assuming reaction after reactant capture) \citep{RN25}. These efficient reactions are therefore important for the chemistry of low temperature environments. 

Therefore, ion-radical reactions are important in determining the chemical composition of the complex systems represented by planetary atmospheres. The reactions of ion-molecule involving open-shell species are generally more complex and unpredictable than those of closed-shell molecules. Recently, charge transfer, which occurs on a very short time scale and over long distances, has emerged as the dominant pathway in ion-radical reactions involving the methyl radical \citep{RN21,RN22}. Significant bond-forming exit channels have also been observed with sufficient internal energy to cause cleavage of the C-H bonds.\par

Laboratory studies in the literature present an analysis of the products formed by UV photon irradiation of relevant gas mixtures of Titan's atmosphere using a variety of sources such as low-pressure mercury and deuterium lamps but with a limited photon energy range from 115 to 254 nm, which does not enable a direct photodissociation of N$_2$  (e.g., \citep{yoon2014role,sebree2014titan, cable2014identification,horst2018laboratory}). 

However, the most interesting energy range is above the nitrogen dissociation threshold ($\sim$ 100 nm) because laboratory experiments in the literature show that the presence of nitrogen increases the complexity of the chemistry in both the gas and solid phases and plays a key role in increasing the efficiency of gas to particle conversion (e.g., \citep{fujii1999analysis,imanaka2007role,trainer2012nitrogen}). One of the most interesting ranges lies above the N$_2$ ionization threshold ($\sim$ 80 nm) leading to N$_2^+$  (with the understanding of its role  still unclear). Imanaka and Smith \citep{imanaka2007role}, followed by Peng et al. \citep{Peng2013},  used synchrotron radiation to show that the formation of neutral unsaturated hydrocarbons is enhanced for wavelengths  $<$ 80 nm, but they were not able to analyze the ions formed at this wavelength. As an alternative to synchrotron radiation, microwave plasma discharge lamps allow to provide selective monochromatic photons in the EUV range. Previous works at room temperature with the irradiation of a Titan's relevant gas mixture (N$_2$/CH$_4$) at 73.6 nm showed the formation of stable N-bearing neutral molecule that is related to the production of N$_2^+$ \citep{RN23}. This work, in continuation of the previous one, focuses on chemical processes in the ionization regimes of N$_2$ and CH$_4$ as in Titan ionosphere, to constrain the N-rich chemistry supporting chemical growth and haze formation at high altitudes in the atmosphere of Titan.

\section{Materials and methods }

\subsection{Experimental simulation of extreme UV photochemistry on Titan}

Our extreme ultraviolet (EUV) source was configured to emit $\sim$10$^{14}$ ph s$^{-1}$ cm$^{-2}$ in total on a 0.78 cm$^2$ beam section at 73.6~nm from a resonant emission of neon triggered by a micro-wave plasma discharge \citep{RN23}. The EUV source is windowless and directly coupled through a movable quarter inch quartz flow tube to the APSIS photochemical reactor  (named for the process of atmospheric photochemistry simulated by surfatron), a stainless steel chamber with a length of d~=~50~cm and a section of 10$^5$~cm$^2$. The experimental setup allowed us to work primarily with EUV photons and without any interaction from other energy sources. The distance between the end of the discharge and the entrance of the reactor was optimized to ensure that electrons and neon metastable atoms are quenched before arriving to the reactive zone \citep{RN32}. Before each experiment, the reactor is pumped down to 10$^{-6}$~mbar. The APSIS reactor is then allowed to flow at room temperature, a partial pressure of reactants of 32 Pa, and at a total gas pressure of 80 Pa. This pressure is ensured by the addition of a 5 sccm (standard cubic centimeters per minute)  of the 95-5\% N$_2$-CH$_4$ reactive gas mixture flux reactive gas mixture flux and a 7.5 sccm neon flux from the EUV source. We note that in the ionosphere of Titan (900–1,500 km), the abundance of CH$_4$ increases progressively from 2\% to 10\% due to the molecular diffusion of lighter species \citep{waite2005ion}. Thus, the 5\% CH$_4$ in the gas mixture used in this work is representative of its average higher amount in the upper atmosphere of Titan. To minimize wall effects, namely, a heterogeneous catalysis on the reactor stainless steel walls, the flow tube used for the injection of the EUV photons was slid into the middle of the reactor. 
In the conditions of this experiment, considering the N$_2$ partial pressure of 30.4 Pa and its absorption cross section of 2.3×10$^-7$ cm$^2$ at 73.6 nm, $\tau$ =1 (one unit of optical depth) occurs at the irradiated column length of ~5.9 cm. Positive ions produced by irradiation of the N$_2$-CH$_4$ gas mixture within the irradiated column are detected in the 1-200 u mass range with a mass resolution of 1 u by quadrupole mass spectrometry. A Hiden Analytical EQP 200 quadrupole mass spectrometer was thus coupled to the reactor through a 100~µm pinhole. Its mass range goes up to 200 u. A turbo pump enables a secondary vacuum of 10$^{-9}$~mbar inside the MS chamber to be reached, in addition to 10$^{-8}$~mbar during ion extraction, avoiding secondary reactions during the transfer of the gas phase through the ion optics of the mass spectrometer. The extractor, where a negative potential is applied, directly extracts ions from the reactor. In all the spectra presented below, counts <10 cps correspond to background noise. Thus, the presented mass spectra thereafter show the direct signal registered without background subtraction. The tip of the QMS is set up 1 cm perpendicularly to the axis of the photon flux (at a 0.6 cm distance from the outer edge of the beam) and a few millimeters after the exit quartz flow tube. Gas sampling is done through a small pinhole (100 microns in diameter). Thus positioned, the mass spectrometer probes the ion content in the first millimeters of the 5.9 cm long irradiated column. The experiments were performed three times to ensure repeatability.

\subsection{0D-Photochemical Model}

A photochemical-0D model was used to reproduce the chemistry obtained in the reactor, interpret the experimental m/z spectra, and further extrapolate the results to Titan’s ionospheric conditions. The model is described in detail in \citep{RN23,Peng}, where it was used both for the interpretation of laboratory experiments and for in situ measurements of positive ions in Titan’s ionosphere by the Cassini-INMS instrument. Recent updates have been made to extend its use to the simulation of photochemistry in exoplanetary atmospheres\citep{Bourgalais20, Bourgalais21}. 
The chemical model comprises 54 photoprocesses, 903 neutral reactions (811 bimolecular and 92 termolecular), and 1941 ion processes (1314 bimolecular and 687 dissociative recombinations), involving 177 neutral species, 190 positive ions of m/z up to 130, and electrons.
The model was run for a time to reach a steady state, about 1 ms, which guarantees the stability of the mole fraction of the main species. This relatively short time has been experimentally supported by the stability of the ion signal of mass spectra recorded after only a few minutes of irradiation in previous studies \citep{Bourgalais20}. The simulated mole fraction of the products is then compared with the experimental data or with the INMS ion mass spectra. Complementary Monte Carlo simulations were performed to assess the uncertainty in the model predictions (500 runs). Subsequently, in the results, the confidence interval used is 2$\sigma$, as in the experimental results. A global rate analysis is performed to identify the key reactions and dominant formation pathways.

\subsection{Theoretical calculations}
{In this work, we study the elemental steps of the reactions of interest by employing
theoretical chemistry. In particular, we obtain: (i) energies of the species of interest such that energy differences of the reaction under consideration are obtained; (ii) rate constants of different elemental processes. 

To this end, we first optimized the geometry of each molecule using density functional theory (DFT) and, notably, the B2PLYP-D3 functional with the aug-cc-pVTZ basis set~\citep{goerigk2011efficient,kendall1992electron}. Then, to have a more accurate
description of the electronic structure (which can be critical due to self-interaction in the current scenario)
we re-optimized them with the highly correlated CCSD(T) method with cc-pVDZ basis set~\citep{purvis1982full,dunning1989gaussian}. Then, using these last geometries, we 
calculated the electronic energies employing a larger basis set, aug-cc-pVQZ, at the same CCSD(T)
level of theory \citep{woon1993gaussian}. Frequency calculations were done using a less computationally demanding method CCSD with the smaller basis set, cc-pVDZ: they are used to add
zero-point energy to all the species and to obtain the projected frequencies along the reaction pathways needed for rate constant calculations.
}
\par
{We calculated the electronic energies of the different species also by
using the Extrapolated Full Configuration Interaction (ExFCI) \citep{garniron2019quantum} with the cc-pVDZ basis set using the geometries obtained at CCSD(T)/cc-pVDZ level of theory. 
ExFCI is used to take into account the strong correlation which is also explicitly 
measured by calculating the  largest amplitude of the double excitations operator, T2.}
As pointed out recently by several of the authors of the present work \citep{giner2018interplay}, the test of strong correlation based on T1 diagnostic is usually not reliable as it mainly expresses orbital relaxation due to correlation effects, which is not a unique signature of strong correlation. Typically, if the largest values of T2 amplitudes are smaller than 0.3-0.4, then it can be safely assumed that the electron-electron correlation is not strong and therefore Coupled Cluster (CC) calculations are reliable. The CCSD, CCSD(T), and B2PLYP calculations are performed using Gaussian 09 package \citep{g09}, while ExFCI is conducted using the Quantum Package \citep{garniron2019quantum}. \par 
{We calculated the rate constants}
using the Langevin model \citep{eyring1936theoretical,gioumousis1958reactions}, for the ion-molecule capture rate, $k_L$, followed by microcanonical variational transition state theory ($\mu$-VT) calculations \citep{garrett1979criterion,hu1991modification,bao2017variational}, to obtain unimolecular dissociation rate constants, $k^{\mu-VT}$. In particular, for N$_2^+$ + CH$_3$ capture providing N$_2$CH$_3^+$ the original Langevin model is used, since the neutral molecule (here CH$_3$) is not polar:

 \begin{equation}
      k_L=2\pi q\sqrt{\frac{\alpha}{M}}
      \label{eq:langevin}
 ,\end{equation}

where $q$ is the charge, $\alpha$ the polarizability of the neutral molecule, and $M$ the reduced mass of the system. The polarizability of the CH$_3$ radical is calculated at both CCSD(T)/aug-cc-pVQZ and B2PLYP/aug-cc-pVTZ levels of theory, providing very similar values (2.39 and 2.38 \AA$^3,$ respectively). These results are close to those calculated with less precise methods previously \citep{jin2003computational,miller1990additivity}. \par 
We note that in this case the capture rate constant is not temperature-dependent. Temperature dependence appears when we also consider the ion-quadrupole interaction, with a $T^{-1/6}$ dependency, which is relevant only at very low temperatures, as also confirmed by laboratory measurements \citep{clary1990fast}.\par 
{We calculated the unimolecular microcanonical fragmentation rate constants of the complex N$_2$CH$_3^+$ via the RRKM theory~\citep{hu1991modification,baer1996unimolecular}: }

 \begin{equation}
      k^{\mu VT}(E)=\sigma \frac{\mathrm{min}_s N^{GT}(E,s)}{\rho_R(E)h}
      \label{eq:mvt}
 ,\end{equation}
   
 where $E$ is the internal energy, $\sigma$ the symmetry factor, $s$ the reaction coordinate, $N^{GT}$ the sum of ro-vibrational states of the generalized transition state along the reaction coordinate, $\rho_R$ the density of ro-vibrational states of the reactant, and $h$ the Planck constant. \par
The transition state is located as the minimum of sum of ro-vibrational states along the reaction pathway, $s$, following its definition in transition state theory. The sum ($N^{GT}$) and density ($\rho_R$) of states are calculated using the direct count algorithm \citep{beyer1973algorithm}, using frequencies obtained from quantum chemistry calculations at CCSD/cc-pVDZ level of theory. The density of states is calculated for the reactant geometry, while $N^{GT}$ for each point along the reaction path, here N-C and H-C distance for reactions (a) and (b), respectively (see Section~\ref{sec:kinetics}).
Notably, for $N^{GT}$, we used the projected vibrational frequencies along the reaction coordinate obtained from a set of relaxed geometries where all the modes, except the dissociating bond, are relaxed. The rotational contribution is directly added from the inertia moments, which are easily calculated from the geometries. This means that for each excess energy, there is a curve of $N^{GT}$ as a function of $s$. Here, we consider an excess energy just above the entrance channel.\par
{We used the Canonical-Variational Transition State Theory (CVT) to obtain
thermal rate constants \citep{garrett1979criterion,hu1991modification,bao2017variational}.} In this case the free energy barrier is located as the maximum of the free energy difference with respect to reactants along the reaction coordinate, $s$, at given temperature, $T$, and inserted into: 

 \begin{equation}
      k^{\mathrm{CVT}}(T)=\frac{k_BT}{h}e^{-\Delta G^\ddagger(T,s)/k_BT}
      \label{CVT}
   ,\end{equation}
   
 where $k_B$ is the Boltzmann constant and $\Delta G^\ddagger$ is the activation free energy for which the maximum along the reaction pathway is used (this quantity is now temperature-dependent). Then, $G$ is calculated as for isolated molecules for each point along the reaction coordinates using the same projected vibrational frequencies used for the $k^{\mu\mathrm{VT}}(E)$. The difference between the reactant, $G,$ and the maximum along $s$ defines the $\Delta G^\ddagger$ at each temperature.\par 
Calculations of sum, density of states, and RRKM rate constants were done with the software developed by Zhu and Hase \citep{Zhu94}, as well as for $G$ using Gaussian09 utilities \citep{g09}, while $\mu$-VT and CVT localization of TS were done with in-house codes.

Finally, we estimated the possibility of a long-range charge transfer from N$_2^+$ to CH$_3$ by evaluating the non-adiabatic coupling. This is done using the geometries obtained from  variational transition state theory calculations. More details on how the non-adiabatic couplings are evaluated are given in the appendix.

\section{Results}
\subsection{Experimental and modeling results}

This work is first conducted by an experimental simulation of the chemical reactivity occurring in Titan’s thermosphere. Molecular nitrogen and methane are irradiated at 73.6 nm, a wavelength that promotes the formation of N$_2^+$ ions since the N$_2$ ionization threshold is at 79.4 nm, and also produces CH$_4^+$ ions (ionization threshold at 98.3 nm). Positive ions produced by irradiation of the N$_2$-CH$_4$ gas mixture are detected by quadrupole mass spectrometry.\par 

\begin{figure}
   \centering
   \includegraphics[width=8cm]{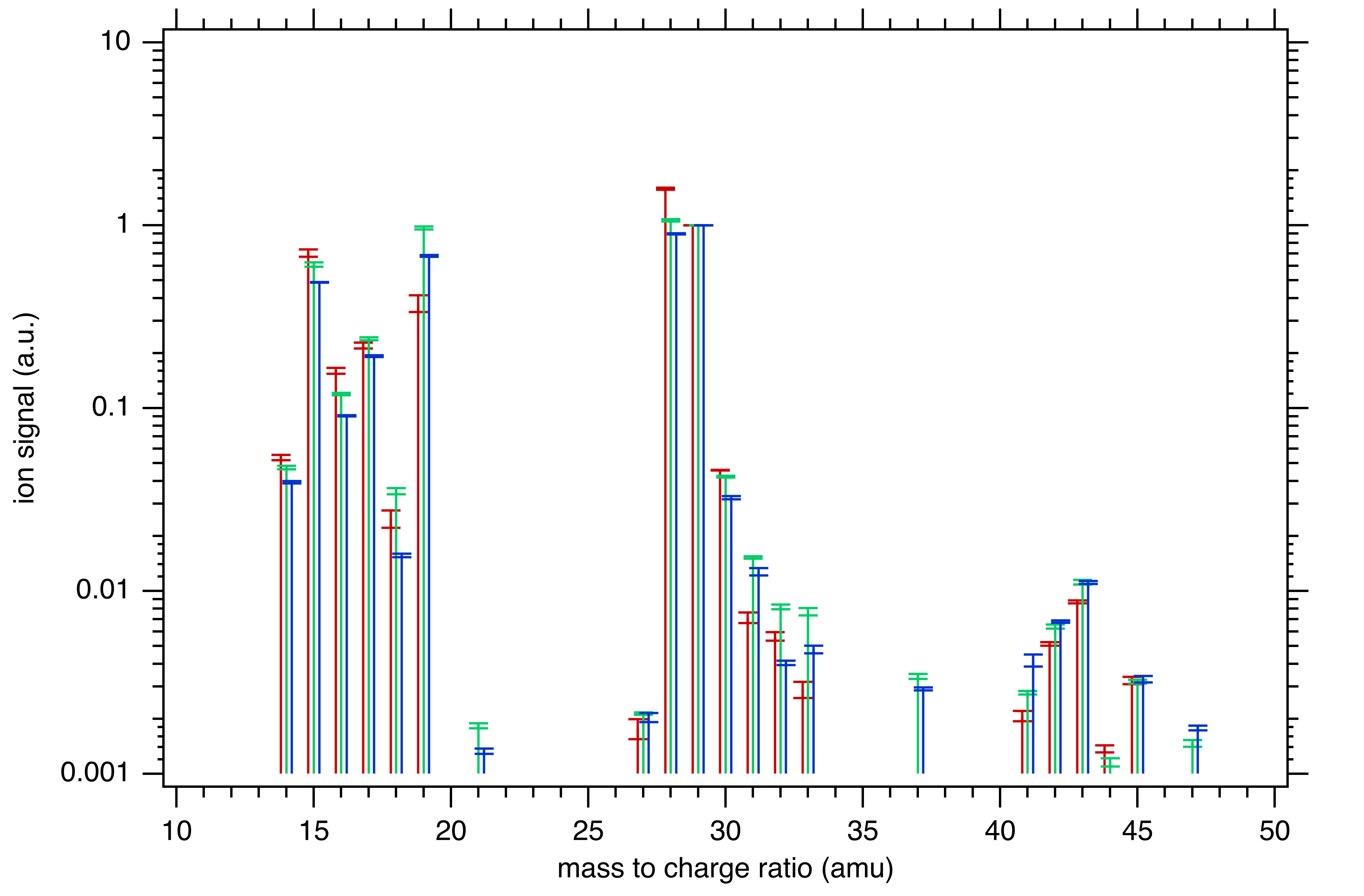}
      \caption{Ions detected for the three repeated experiments of irradiation of a N$_2$-CH$_4$ 95-5 \% gas mixture with photons at 73.6 nm and a reactant partial pressure of 32 Pa. The three experiments are reported with three distinct colors and the mass spectra are slightly shifted for clarity. Uncertainties are reported on each ion for each experiment, corresponding to twice the standard deviation of the ion signal variability, obtained by measuring a few tens of spectra for each experiment.}
         \label{Fig1}
   \end{figure}

The results are given in Figure \ref{Fig1}. The intensities are normalized by the intensity of the most abundant ion, C$_2$H$_5^+$ at $m/z$~29. The uncertainty is reported on each ion for each of the three experiments. It corresponds to twice the standard deviation of the variability of the ion signals, based on the acquisition of a few tens of spectra for each experiment. 

\subsubsection{Light positive ions ($m/z$ $\leq$ {29})}
Previous work has been carried out under similar conditions \citep{RN23} but with a lower pressure by almost two orders of magnitude (total pressure of 1 Pa vs 80 Pa here). In the previous case, only light ions with $m/z$ $\leq$ {29} could be observed due to the sensitivity limit of our mass spectrometer. The interest of the higher pressure conditions is revealed by the detection of ions at masses larger than $m/z$ 29.

\begin{figure}
\centering
 \includegraphics[width=8cm]{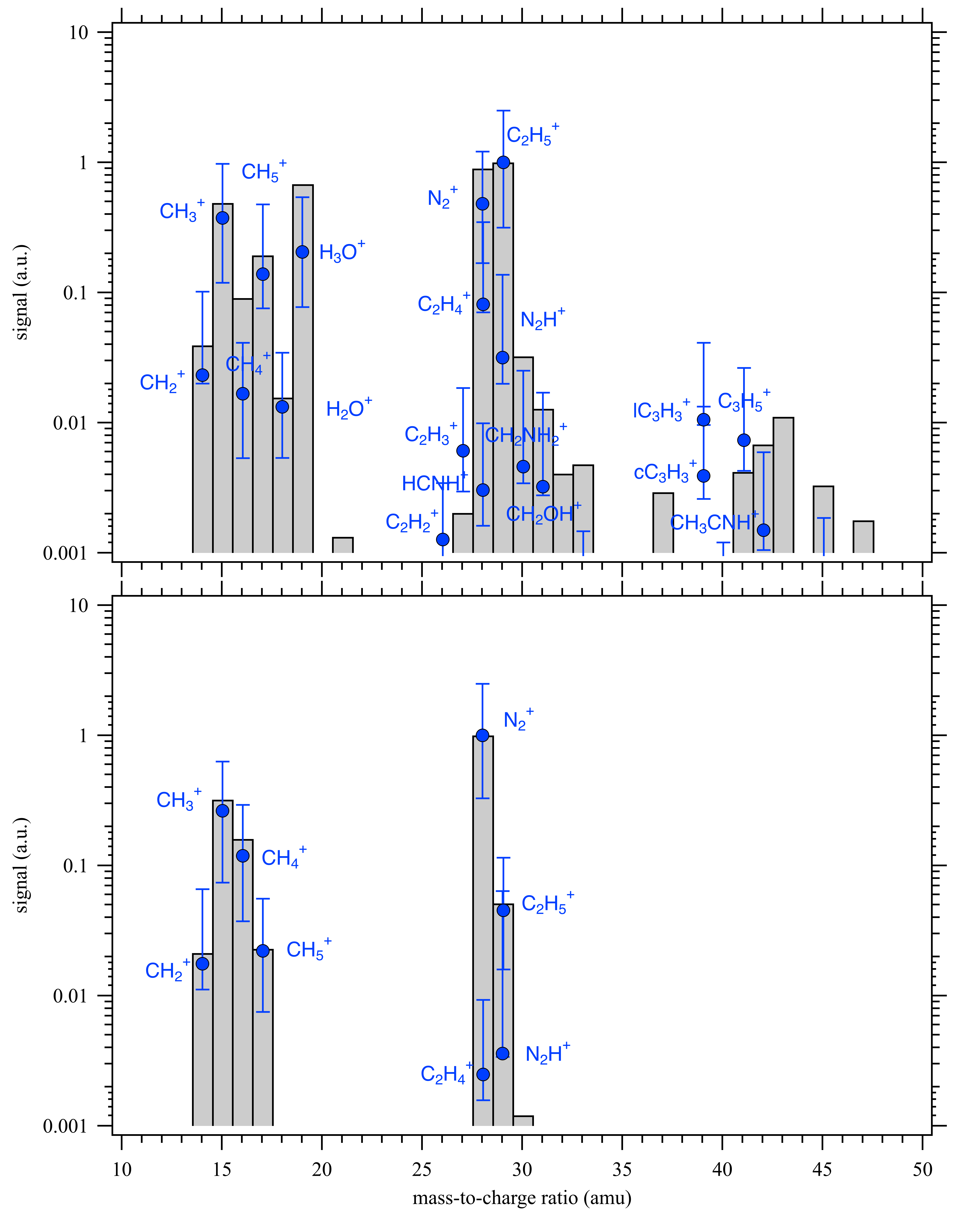} 
       \caption{
        Ion mass spectra obtained with the same experimental conditions except the pressure (lower panel) at 1 Pa in \citep{RN23} and (upper panel) at 80 Pa in the present work. Model predictions are superimposed for each case. 
       }
         \label{Fig2bis}
\end{figure}

However, we have to control that this pressure increase in the reactor does not change drastically the general chemical network at work in the photoreactor. For this purpose, we compare in Figure \ref{Fig2bis} the ion mass spectra observed in two experiments at 1 and 80 Pa and the results of their respective chemical modeling. 
First the chemical modeling results obtained are in full agreement with the experimental data for the ions at $m/z$ $\leq$ {29} for both experimental data, confirming that the chemistry is well constrained for the light ions in the considered [1-80] Pa total pressure range. The modeling results thus enable us to identify the ions observed experimentally as well as their formation and loss processes.
The first noticeable difference between the two experiments is the presence of water ions H$_2$O$^+$ and H$_3$O$^+$ (at $m/z$ 18 and 19, respectively) in the reactor for the experiment at 80 Pa. This water contamination, even if stable in the present dataset, could be problematic if it were coupled with the other ions. No other O-bearing ions than H$_2$O$^+$ and H$_3$O$^+$ is observed, suggesting that the water ions are independent from the rest of the N$_2$-CH$_4$ ionic chemistry generated in the photoreactor. To confirm this hypothesis, a complementary numerical simulation is led with the same parameters as for the simulation at 80 Pa, but water-free. The ions (others than H$_2$O$^+$ and H$_3$O$^+$) and their relative densities remain unchanged, confirming that the N$_2$-CH$_4$ ionic chemistry is not affected by the presence of the water contamination.  
Then the other main difference between the two sets of experiments is the relative intensities of the ions. The ions are qualitatively the same, but at 1 Pa total pressure, the highest densities are observed for CH$_3^+$ and N$_2^+$, which are primary ions formed by the photolysis of CH$_4$ and N$_2$, respectively. At 80 Pa, the main ions become CH$_5^+$ and C$_2$H$_5^+$. A flux analysis shows that these ions are mainly produced by the following ion-molecule reactions:\\
N$_2$H$^+$ + CH$_4$~$\longrightarrow$ CH$_5^+$ + N$_2,$ \\
CH$_3^+$ + CH$_4$~$\longrightarrow$ C$_2$H$_5^+$ + H$_2.$ \\
This analysis confirms that the use of a higher pressure in the present case not only enables the enhancement of the detection efficiency by the mass spectrometer, but also to probe the N$_2$-CH$_4$ ionic chemical network in a more advanced stage compared to the experiment at 1 Pa.
In what follows, we focus on understanding the formation of these new heavier ions observed at $m/z$>29.\par

\subsubsection{Heavier positive ions ($m/z$ >29) and isotopic labeling experiments}
 
\begin{figure}
\centering
 \includegraphics[width=9cm]{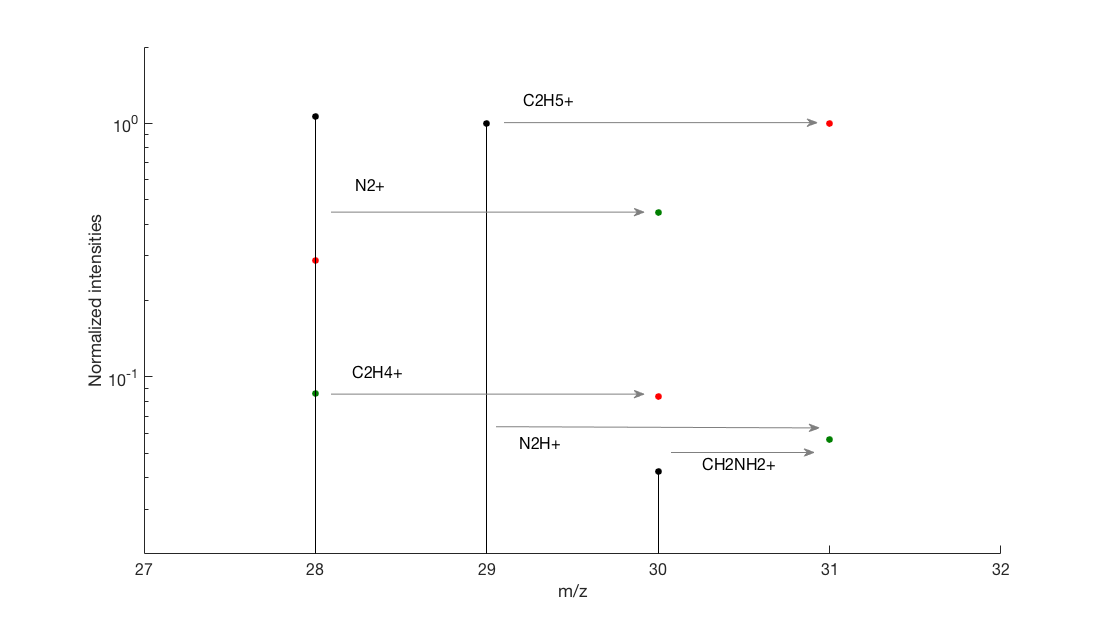} 
 \includegraphics[width=9cm]{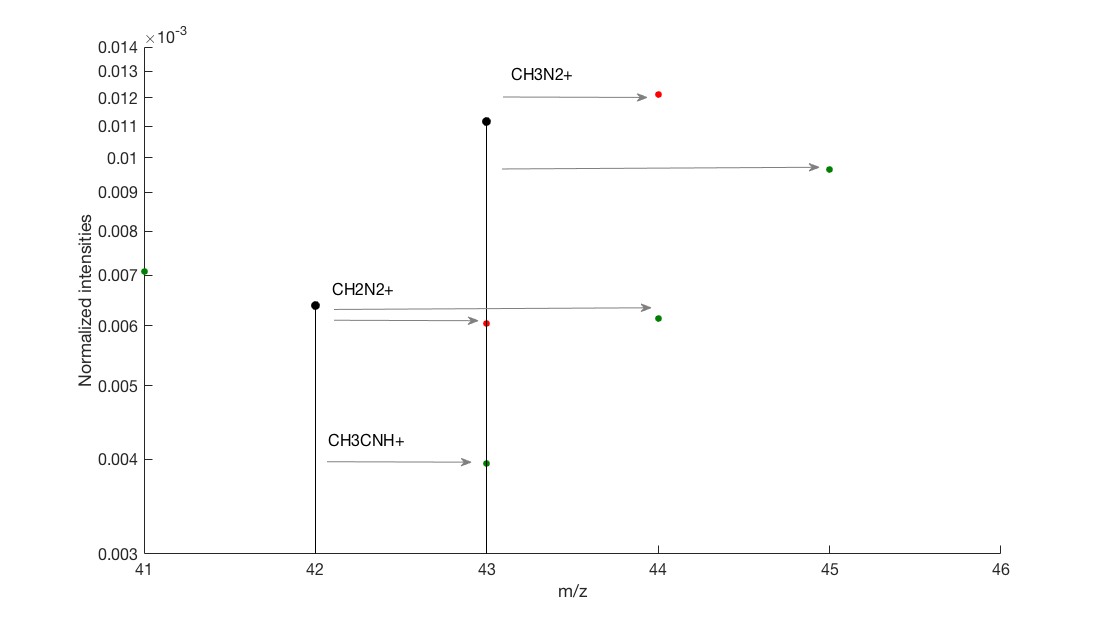} 
       \caption{(black) Positive ion spectra of the three regular experiments normalized to the peak at $m/z$ 29, and compared to the same experiments lead with isotopic gases: (red) $^{13}$CH$_4$, and (green) with $^{15}$N$^{15}$N. The upper panel corresponds to the mass range 27-32 u (C2 block), whereas the lower panel corresponds to the mass range 41-46 u (C3 block).
       }
         \label{Fig2}
\end{figure}

The mass resolution of our mass spectrometer precludes our ability to distinguish isobaric ions. In addition to the modeling work, we also performed experiments using isotopic gases to identify the ions detected in Figure \ref{Fig1}: one experiment combining regular N$_2$ and $^{13}$CH$_4$, and one experiment combining $^{15}$N$^{15}$N and regular CH$_4$. A shift of one mass unit for an ion produced with the gas mixture containing $^{13}$CH$_4$ indicates that the ion has one carbon atom, whereas a shift of one mass unit for an ion produced with the gas mixture containing $^{15}$N$^{15}$N shows that the ion contains one nitrogen atom. Compared to the reference case, the combined shifts observed with both labeled experiments enable the explicit identification of all C$_x$H$_y$N$_z^+$ ions. The results are given in Figure \ref{Fig2}: the upper panel corresponds to the mass range 27-32 u and is called block C2 (two heavy atoms, C or N, per ion), whereas the lower panel corresponds to the mass range 41-46 u and is called block C3 (three heavy atoms, C or N, per ion). The identification process is summarized in Table \ref{Tab1}.


\begin{table*}
\caption[]{Identification of the ions composing the reference mass spectrum thanks to the mass shifts observed with the $^{15}$N and $^{13}$C isotopic labeled experiments.}
\label{Tab1}
\centering
\begin{tabular}{ccccl}

            \hline
            \noalign{\smallskip}
            Block      &  $m/z$ & Shift with $^{15}$N &  Shift with $^{13}$C & Ion \\
            \noalign{\smallskip}
            \hline
            \hline
            \noalign{\smallskip}
            C2 & 28 & +2 & 0 & N$_2^+$   \\
               & 28 & 0 & +2 & C$_2$H$_4^+$   \\
               & 29 & 0 & +2 & C$_2$H$_5^+$   \\
               & 29 & +2 & 0 & N$_2$H$^+$   \\
               & 30 & +1 & +1 & CH$_2$NH$_2^+$   \\
            \hline
            C3 & 42 & +1 & +2 & CH$_3$CNH$^+$  \\
               & 42 & +2 & +1 & CH$_2$N$_2^+$   \\
               & 43 & +2 & +1 & CH$_3$N$_2^+$   \\
            \hline
\end{tabular}
\end{table*}

The C2 block is based on three main signatures in the reference spectrum at $m/z$ 28, 29, and 30. The isotopic labeling highlights that the first peak at $m/z$ 28 is actually the sum of two major ion contributions, N$_2^+$ and C$_2$H$_4^+$. Then, the signal at $m/z$ 29 corresponds almost exclusively to the C$_2$H$_5^+$ ion, but the increase of the intensity at $m/z$ 31 with the $^15$N isotopically labeled experiment also highlights a minor contribution of N$_2$H$^+$. Finally, the peak at $m/z$ 30 corresponds to the protonated methanimine CH$_2$NH$_2^+$: it is illustrated by a shift of +1 with the $^{15}$N isotopically labeled experiment. For this ion, the shift of +1 with the $^{13}$C isotopically labeled experiment cannot be distinguished from the large contribution at $m/z$ 31 also provided by C$_2$H$_5^+$. C$_2$H$_6^+$ is actually discarded at $m/z$ 30 because no detectable increase is observed at $m/z$ 32 with the $^{13}$C labeling experiment. So if present, C$_2$H$_6^+$ is negligible compared to CH$_2$NH$_2^+$ at $m/z$ 30.\par 
Block C2 thus first confirms the abundant formation of the primary ion N$_2^+$ in our photochemical experiment using an irradiation wavelength of 73.6 nm. The intense signature of C$_2$H$_5^+$ is also consistent with the methane ion-neutral reactivity, amplified by charge-transfer reactions with N$_2^+$, as discussed in previous EUV-photochemical studies \citep{RN25,RN26}. More importantly, these ions based on pure nitrogen and methane coexist with the protonated methanimine CH$_2$NH$_2^+$. The C2 block is therefore also representative of a first coupling between nitrogen and methane by ion chemistry, with a balanced N/C ratio of 1 in the protonated methanimine ion. The ion HCNH$^+$ is, however, minor if present in this study, consistent with the modest advancement stage of the chemical network in the first mm of the irradiated column, where the mass spectrometer is positioned. \par

Block C3 is surprisingly mainly composed of two ion peaks at $m/z$ 42 and 43. The photochemical modeling (Figure \ref{Fig2bis}) indeed predicts no significant ion mole fraction at $m/z$ 42 and 43, but possible weak productions at $m/z$ 39 and 41 for C$_3$H$_3^+$ and C$_3$H$_5^+$ ions respectively. The low abundances of C$_3$H$_3^+$ and C$_3$H$_5^+$ likely explain their none detection by the mass spectrometer. However the significant higher productions of the ions at $m/z$ 42 and 43 are not predicted, pointing out a lack in the photochemical model. At $m/z$ 42, protonated acetonitrile CH$_3$CNH$^+$ contributes slightly as shown by the small contribution to a shift of +1 u observed in Figure \ref{Fig2} with the $^{15}$N isotopically labeled experiment, but it is far from dominant. The major contribution at $m/z$ 42 has a shift of +1 u with  $^{13}$C and +2 with the  $^{15}$N isotopic labeling: it thus corresponds to the CH$_2$N$_2^+$ ion. Similarly, the ion peak at $m/z$ 43 is identified as CH$_3$N$_2^+$ and also contains two nitrogen atoms for one carbon atom in the structure. Both ions CH$_2$N$_2^+$ and CH$_3$N$_2^+$ have a N/C ratio of 2 and are absent from usual databases in astrochemistry \citep{KIDA,UMIST}. Their formation is intriguing. No mechanism identified in gas-phase chemical databases explains the formation of these highly nitrogenated ions, except a marginal reaction C$_2$H$_7^+$ + H$_2$CN$_2$ ${\longrightarrow}$ ~CH$_3$N$_2^+$ + C$_6$H$_6$, reported in \citep{Anicich2003}. \par
To explain their formation, we have to infer an ion-molecule reaction that could explain the formation of organic positive ions with three heavy atoms, including two nitrogen atoms. The experiment was conducted with an EUV radiation source at a wavelength of 73.6~nm and this wavelength triggers the photoionization of N$_2$ and CH$_4$ molecules: \\

N$_2$ $\overset{73.6~nm}{\longrightarrow}$~N$_2^+$ + e$^-,$

CH$_4$ $\overset{73.6~nm}{\longrightarrow}$~CH$_4^+$ + e$^-.$ \\
 
As previously studied \citep{RN25}, CH$_4^+$ is moreover rapidly converted into CH$_5^+$ and CH$_3$ from the rapid charge transfer reaction with methane:\\
   
 CH$_4^+$ + CH$_4^+$ $\longrightarrow$~CH$_5^+$ + CH$_3.$ \\
   
As a consequence, CH$_3$ radicals are the most abundant C-containing neutral after methane in our experimental conditions. The presence of two nitrogen atoms in the detected diazo-ions informs the reactions to be explored, either towards N$_2^+$ with an abundant C-containing neutral or towards N$_2$ reactions with an abundant C-containing ion. Given the abundances of the ions and radicals in the reactive medium, the possible reactions that could lead to diazo-ions might involve N$_2$ and CH$_4^+$ and CH$_3^+$ on the one side, and N$_2^+$ with CH$_3$ on the other side. The N$_2^+$ + CH$_4$ reaction should be discarded in principle because it has been extensively studied in the past and leads neither to CH$_2$N$_2^+$ nor CH$_3$N$_2^+$ ions \citep{RN27}. The following section provides the first calculations on these reactions to explore the possibility that different reactions may explain the N$_2$-rich ions observed in the experiment.
\\

\subsection{Theoretical calculations}
\subsubsection{Energetic of diazo formation reactions}

We have first considered the energetic of the three possible reactions (details on the energies and comparison between different methods are given in the Appendix):

\begin{enumerate}[(1)]
    \item N$_2^+$ + CH$_3$,
    \item N$_2$ + CH$_3^+$,
    \item N$_2$ + CH$_4^+$.
\end{enumerate}
Reaction (1) can lead to the association product, N$_2$CH$_3^+$, or by producing atomic H to N$_2$CH$_2^+$. Both of these reactions are highly exothermic as detailed 
in Table~\ref{table:reactenergy}, and they will be discussed in more details in next subsection. We can anticipate that there are no energy barriers to reach the different possible products. \par 
Reaction (2) can form the N$_2$CH$_3^+$ association product, while the formation of N$_2$CH$_2^+$ + H is endothermic (by more than 60 kcal/mol, see Table~\ref{tab:react2}) and thus not possible under Titan’s condition. The stable species N$_2$CH$_3^+$ can be stabilized in higher pressure conditions (and likely some can be in our experimental conditions) but is not stable in the low-pressure conditions of Titan’s atmosphere because it has enough energy to reverse to its reactants.\par
Finally, reaction (3) can form both N$_2$CH$_3^+$ and N$_2$CH$_2^+$ products (by losing neutral H and H$_2$, respectively), which are slightly exothermic (see Table~\ref{tab:react2}). The reaction should proceed by forming first a complex between N$_2$ and CH$_4^+$, which was identified to be a (N$_2$:H:CH$_3$)$^+$ species where H is between N and C atom. The stable N$_2$CH$_3^+$ and N$_2$CH$_2^+$ species have a N-C bond, so this H must be removed from the position it has in this intermediate. This can occur via the formation of a stable intermediate (HN$_2$:CH$_3$)$^+$, but in this state the system has to pass through a transition state which is higher in energy than the entrance channel. We note that none of this reaction product was observed in the experimental study reported by Dutuit et al. \citep{RN25}, in agreement with the conclusions of our calculations. In summary, reaction (1) seems to be the most probable to be responsible of the formation of diazo species in Titan’s atmosphere conditions. We thus went on to study it in greater detail.\par 

\subsubsection{Geometries and energies of the N$_2^+$ + CH$_3$ reaction}

The N$_2^+$ + CH$_3$ reaction was thus investigated using highly correlated methods with an extended basis set for the calculation of the energies of the relevant structures. The rate constants were estimated using variational transition state theory (VTST) on such highly correlated results for unimolecular dissociation of the most stable products and the Langevin model for the formation of N$_2$CH$_3^+$. The robustness of the method was verified by comparison with state-of-the-art and near exact multi-reference calculations in modest basis in order to quantify the level of strong correlation. \par
\begin{figure}
\centering
 \includegraphics[width=8cm]{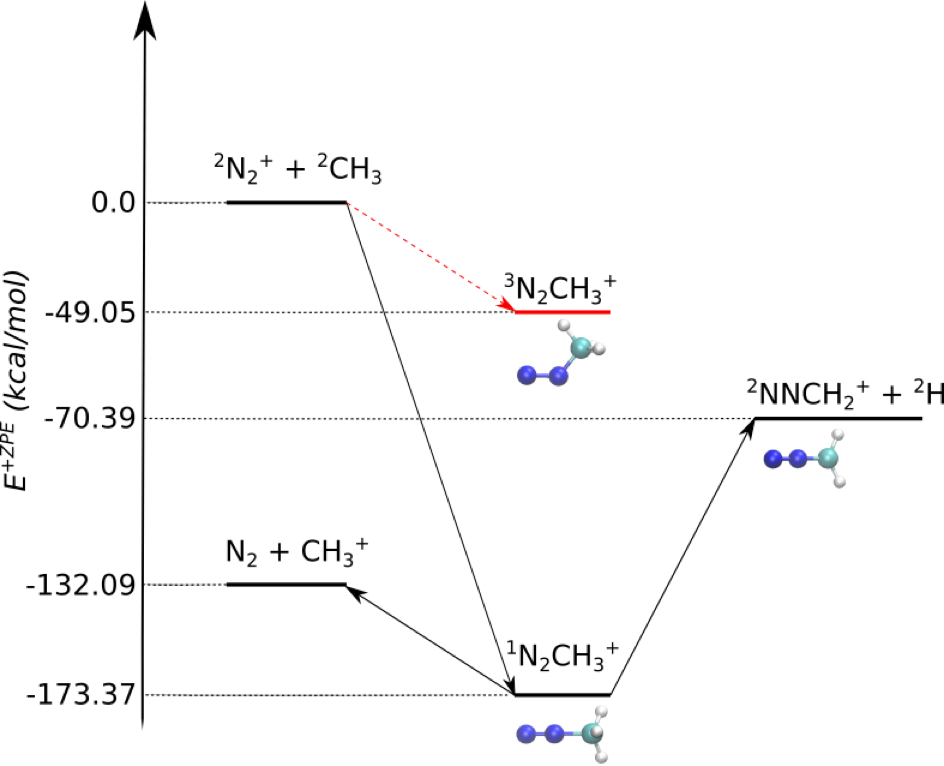} 
       \caption{Schematic reaction pathways relevant to the $^2$N$_2^+$ + CH$_3$ reaction. In red, we report the triplet state, which is largely unlikely. The relative energies (electronic + ZPE) reported are those obtained from the CCSD(T)/aug-cc-pVQZ//CCSD(T)/cc-pVDZ calculations. For $^2$NNCH$_2^+$, we report the optimized structure of the most stable isomer.}
         \label{Fig3}
\end{figure}

The N$_2^+$ + CH$_3$ reaction can lead to several exit channels with different products, which we have classified as follows. They are represented with their energies in Figure \ref{Fig3} (where zero is set to the entrance channel, N$_2^+$ + CH$_3$):

Ion-molecule complex: We obtained as stable species the N$_2$CH$_3^+$ ion in both singlet and triplet states. The singlet is a linear structure while the triplet is bent. The geometries are shown in Figure \ref{Fig3}. The triplet state is clearly an excited state and is unlikely to be populated. The singlet state $^1$N$_2$CH$_3^+$ is much lower in energy and is even lower than the N$_2$ + CH$_3^+$ product.

The N$_2$ + CH$_3^+$ product: This exit pathway is more stable than the entrance channel of 133 kcal/mol and it can therefore be considered that, if after the collision, N$_2$ and CH$_3$ are scattered, the charge will jump to the methyl radical. This product can be obtained also from long-range charge transfer from N$_2^+$ to CH$_3$ directly without passing through an intermediate complex. We estimated the non-adiabatic couplings between the two states with different charge localization (on N$_2$ and on CH$_3$) and, as detailed in the appendix, these values are zero or very
small, such that this process can be considered negligible at a first approximation. 

Reactive scattering products $^2$N$_2$CH$_2^+$ + $^2$H: Several structures of the same stoichiometry are possible and are shown in the Appendix (Figure~\ref{FigS1}), all of which are more stable than the entrance channel and can therefore be formed. Two structures are the most stable, $^2$N$_2$CH$_2^+$ and $^2$HNCNH$^+$. The first can be easily obtained from N$_2$CH$_3^+$ by losing a H-atom, while the second requires rearrangement followed by the breaking of the strong N-N bond. We can thus consider only the first species for kinetic calculations and understanding the competition with the N$_2$ + CH$_3^+$ exit channel. Other asymptotes might also contribute, reinforcing the formation of diazo-ions.

Reactive scattering products N$_2$CH$_2$ + H$^+$: In this case, several structures for neutral N$_2$CH$_2$ are also possible. However, they all have a higher energy than the entrance channel and therefore are not taken into account in the following.

\subsubsection{Kinetics}\label{sec:kinetics}

Based on the structures and energies of stationary points, we can estimate the rate constants of the different processes. \par The first step is the formation of CH$_3$N$_2^+$ ion from the ion-molecule reaction N$_2^+$ + CH$_3$. A capture rate constant of 1.14~$\times~$10$^{-9}$ cm$^3$ s$^{-1}$ is obtained using the Langevin model (see Eq.~\ref{eq:langevin}).\par 
We then focused our kinetic study on the two most plausible reactions that the CH$_3$N$_2^+$ ion (in the single state) can undergo:
\begin{enumerate}[(a)]
    \item CH$_3$N$_2^+$~$\longrightarrow$~CH$_3^+$ + N$_2$,
    \item CH$_3$N$_2^+$~$\longrightarrow$~N$_2$CH$_2^+$ + H.
\end{enumerate}

In this case, we consider the C-N and C-H distances as reaction coordinates for (a) and (b), respectively. The energy profile shows that the kinetics is governed by loose transition states (TSs) and in Figure~\ref{FigS2} of the appendix, we report the internal energy and the sum of rovibrational states ($N^{GT}$) as a function of the reaction coordinates for an internal energy value of 174 kcal/mol, which is just above the entrance channel. The corresponding microcanonical rate constants are 5.29$\times10^{13}$ and 5.12$\times10^{11}$ s$^{-1}$, from which we can estimate that the branching ratio of N$_2$CH$_2^+$ product is 1\% and that the global microcanonical rate constants leading to products (a) and (b) are 1.13$\times10^{-9}$ and 1.10$\times10^{-11}$ cm$^{-3}$ s$^{-1}$, respectively. \par
This means that in a purely collisional regime (no radiative or collisional stabilization of the complex), one percent of the first collisional products should be N$_2$CH$_2^+$. The radiative stabilization rate constant was actually estimated to be 55 s$^{-1}$, which is much slower than the competitive unimolecular fragmentations. However, the presence of a gas can stabilize the N$_2$CH$_3^+$ ion product. If the system is fully thermalized, then the canonical kinetics should be considered, and the free energy barriers and rate constants must be calculated using the CVT method (Eq.~\ref{CVT}). We obtained free energy barriers at 300 K of 31 and 110 kcal/mol with the corresponding very slow rate constants of 1.24$\times10^{-10}$ and 1.72$\times10^{-68}$ s$^{-1}$. We note that the reactions are characterized by loose TSs and thus the structures are not saddle points of the potential energy but maxima in the free energy profile, which are temperature-dependent, and they are mainly characterized by N-C and H-C distances, which are, at 300 K, 3.58 and 3.10 \AA  for reaction (a) and (b), respectively.\par

\subsubsection{Limit case theoretical scenarios}

The N$_2^+$ ions created by photoionization when it reacts with CH$_3$ radicals will either lead to the stabilization of the methyldiazonium adduct-ion, N$_2$CH$_3^+$, or dissociate to lead to the formation of other ions. Based on the potential energy surfaces and kinetic calculations, the following picture should be considered: \par

First, if the energy is not dissipated, the methyldiazonium ion dissociates to lead to the products N$_2$ and CH$_3^+$ by charge transfer, or lead to a diazo compound N$_2$CH$_2^+$ via the loss of a H-atom. This last reaction is also barrier-less, and although it has an energy  higher than the charge transfer products channel, it is nevertheless possible and expected to be significant with an estimated branching ratio of about 1\%. This is the limit case in which no thermalization occurs and with energy slightly higher than the entrance channel. As we discuss in the next section, this apparently low percentage can be relevant to the formation of N$_2$CH$_2^+$ ion under Titan’s atmosphere conditions.

Second, if there is thermalization, it means that the ion N$_2$CH$_3^+$ can be formed and stabilized. The free energy barrier to reach N$_2$ + CH$_3^+$ products is relatively high at 300 K (about 31 kcal/mol), while the other products are unreachable, with a barrier of 110 kcal/mol. 

It should be noted that these scenarios are indeed two limiting theoretical scenarios where assumptions are made about kinetic or thermodynamic regimes. It has been shown that a bimolecular reaction can be largely affected by non-statistical (or dynamical) effects that may play an important role and cannot be described with simple kinetic or thermodynamic pictures \citep{RN28,RN29}. To perform such studies, one will need a way of calculating on the fly the interaction potential with methods that are faster than those used here (likely with some density functional theory). Qualitatively, it can be expected that, as observed recently, the high-energy product will be more populated than reported in the kinetic study \citep{RN30}. Furthermore, the reactants can have more energy (due for example to photoionization) and thus change the branching ratio. \par

\section{
\textbf{Impact on Titan's atmosphere}
}

The most important result given by the theoretical calculations is to reveal that the adduct diazo-ion N$_2$CH$_3^+$ is actually formed after reaction of N$_2^+$ with the CH$_3$ radical. N$_2$CH$_3^+$ is energetically even more stable than the “N$_2$ + CH$_3^+$” product of the charge transfer pathway.

\subsection{Estimation of the partial rate constant of the reaction from the experimental data}

This important theoretical result legitimates to introduce a new reaction pathway in the photochemical model simulating our experiment:\\
(1) N$_2^+$ + CH$_3$ $\overset{k_1}{\longrightarrow}$ CH$_3$N$_2^+$ ; $k_1$ (cm$^3$ s$^{-1}$)\\

In this case, the value of this ion-neutral reaction is interestingly inferred from our experimental data and can be confronted to the theoretical prediction. 
As a simple first approximation, we consider that the diazo cation CH$_3$N$_2^+$ is produced by reaction (1) and consumed by dissociative recombination with free electrons (DR). The stationary state concentration of the CH$_3$N$_2^+$ diazo cation that we observe experimentally is thus controlled by the following equilibrium:

\begin{equation}
    \frac{\mathrm{d}n\mathrm{CH}_3\mathrm{N}_2^+}{\mathrm{d}t} = k_1\times n \mathrm{CH}_3 \times n \mathrm{N}_2^+ - k_{DR} \times n \mathrm{e}^- \times n \mathrm{CH}_3\mathrm{N}_2^+=0, 
\end{equation}

leading to the value for the partial rate constant k$_1$:

\begin{equation}
    k_1= k_{DR}\times \frac{n \mathrm{e}^- \times n \mathrm{CH}_3\mathrm{N}_2^+}{n \mathrm{CH}_3 \times n \mathrm{N}_2^+} 
.\end{equation}

The experimental data provide a value of the ratio 
\begin{math} n\mathrm{CH}_3\mathrm{N}_2^+ \end{math}/\begin{math} n\mathrm{N}_2^+ \end{math} of 2 $\times$ 10$^{-2}$, whereas the photochemical modeling of the experiment complementary provides the molar fractions of CH$_3$ and electrons, 1.1 $\times$ 10$^{-4}$ and 2.2 $\times$ 10$^{-5}$, respectively. 

The unknown dissociative recombination rate constant of the CH$_3$N$_2^+$ ion is derived from the review work of \citep{Plessis2010}. A confidence interval of DR rate constants for ions with more than three heavy atoms is given in the range  [ 5 $\times$ 10$^{-7}$ , ~3 $\times$ 10$^{-6}$ ] cm$^3$.s$^{-1}$,  with a mean value of 9.5 $\times$ 10$^{-7}$ cm$^3$.s$^{-1}$.

In this case, we obtain a first experimental estimation of k$_1$ = 3.8 $\times$ 10$^{-9}$ cm$^3$.s$^{-1}$ with large uncertainties supported by the absence of knowledge on k$_{DR}$. Such an estimated high value is in the same order of magnitude as the capture rate constant calculated using the Langevin model. This indicates that the diazo cations CH$_3$N$_2^+$ is efficiently formed and stabilized in our experimental conditions. 

\subsection{Extrapolation to Titan’s ionospheric conditions }

At about 1000 km in altitude, where the ions were measured by the Cassini-INMS instrument in the Titan ionosphere, the pressure is much lower, at about 10$^{-6}$ mbar. Under these conditions, the thermalization of the diazo-adduct CH$_3$N$_2^+$ is likely less efficient than in our experiment. We can predict the impact of the new reaction N$_2^+$ + CH$_3$ according to the two extreme scenarios.\\
In the fully thermalized scenario, the reaction remains: \\
(1) N$_2^+$ + CH$_3$ $\overset{k_1}{\longrightarrow}$ CH$_3$N$_2^+$ ; $k_1$ $\sim$ 1.1×10$^{-9}$ cm$^3$ s$^{-1}$, 
\\
with k$_1$ predicted by the Langevin model.
\\
\\
However, if the energy is not dissipated, CH$_3$N$_2^+$ will be converted into CH$_2$N$_2^+$ with a 1$\%$ conversion yield, and the reaction becomes:\\
(2) N$_2^+$ + CH$_3$ $\overset{k_2}{\longrightarrow}$ CH$_2$N$_2^+$ + H ; $k_2$ $\sim$ 1.1×10$^{-11}$ cm$^3$ s$^{-1}$.\\

The intermediate state CH$_3$N$_2^+$ could also be stabilized by radiative emission in Titan’s low pressure ionospheric conditions, and by thermalization in the higher 1 mbar pressure condition of our experiment:\\ 

(3) CH$_3$N$_2^+$ $\overset{k_4}{\longrightarrow}$ CH$_3$N$_2^+$, 

(4) CH$_3$N$_2^+$ + M $\overset{k_5}{\longrightarrow}$ CH$_3$N$_2^+$ + M,

with M the termolecular body in the given environment. 

Radiative stabilization rate constant of reaction (3) are estimated using the method described by Herbst \citep{RN31} and is found to be 55.45 s$^{-1}$ at the entrance channel energy, so it can be neglected. Reaction (4) becomes also negligible in the low-pressure condition of Titan’s upper atmosphere ($\sim$25 s$^{-1}$ at the nanobar pressure). \par

\begin{table*}
\caption[]{Relative mole fractions of CH$_3$N$_2^+$ and CH$_2$N$_2^+$ compared to the most abundant ion in Titan's ionosphere HCNH$^+$.}
\label{Tab2}
\centering
\begin{tabular}{|l|c|c|}
  \hline
  Scenario & CH$_3$N$_2^+$ & CH$_2$N$_2^+$ \\
  \hline
  Thermalization & $\sim$ 1.1×10$^{-2}$ & $\sim$ 0   \\
  No energy dissipation   & $\sim$ 0 & $\sim$ 1.1×10$^{-4}$ \\
  \hline
\end{tabular}
\end{table*}

With our photochemical model we were able to consider these two extreme cases with environmental conditions consistent with Titan’s ionosphere at 1100 km of altitude. The results are reported in Table \ref{Tab2}, where the diazo-ions mole fraction are compared to the mole fraction of the most abundant ion at this altitude, HCNH$^+$ ion. In both cases, the predictions for diazo-ions mole fraction range between 10$^{-2}$ and 10$^{-4}$ compared to the most abundant ion HCNH$^+$. Even the lowest estimation at 10$^{-4}$ is in the same order of magnitude as other important ions like protonated methanamine CH$_3$NH$_3^+$ and the CNC$^+$ ions predicted in Titan’s ionosphere \citep{RN30}. Diazo-ions are therefore found to contribute significantly to the ion budget in Titan’s ionosphere through the reaction N$_2^+$ + CH$_3$. 
\\

\section{Conclusion}

In conclusion, our work shows the existence of new reactions that have not been taken into account in current models to date, which are the first to explain the formation of diazo-ions in agreement with our experimental discovery. We predict the formation of significant new ions, CH$_3$N$_2^+$ and CH$_2$N$_2^+$, with a high impact on Titan chemistry. However, in the framework of this study, we focused on pathways leading to the formation of the diazo cations: it would be valuable in the future to complete the description of all the pathways triggered by the interaction between N$_2^+$ ion and CH$_3$ radical. Moreover this efficient mechanism does not preclude possible additional contributions by other unknown reactions that would also be worth to study in the future. The impact of such additional reactions would even more amplify the important formation of diazo-ions in Titan’s atmosphere revealed in the present work.\par
The formation of diazo ions in Titan’s ionosphere opens up a new branch of knowledge in Titan’s ionospheric chemistry. Also, more generally, this study opens up new perspectives on the importance on studying experimentally and theoretically ion-radical reactions in astrochemistry. Indeed, we showed that the reactions of ions with free radicals can contribute to the richness of ion-neutral chemistry in Titan's atmosphere by forming diazo compounds. In conjunction with charge transfer reactions, which are particularly important in this class of reactions due to the low ionization potential of free radicals, bond-forming reactions of the type reported here, even though they are minority pathways, will play a major role in the propagation of N-rich chemistry. Further experimental studies are needed, supported by ab initio calculations, to learn more about the chemical reactivity that can occur between ions and hydrocarbon-based free radicals in Titan's atmosphere. Diazo compounds are, moreover, very reactive. With their N/C ratio of 2, they are able to incorporate nitrogen into the chemical organic chains of Titan’s atmosphere with unprecedented efficiency. Chemical analysis has actually shown a consistent presence of diazo signatures in laboratory analogues of Titan’s atmospheric aerosols \citep{RN33}. The reaction between N$_2^+$ and CH$_3$ in Titan’s upper atmosphere and the formation of diazo-ions may be a piece of the puzzle to explain the nitrogen insertion pathways in photochemical organic aerosols.

\begin{acknowledgements}
NC and JB thank the European Research Council for funding via the ERC PrimChem and ERC OxyPlanets projects (grant agreements No. 636829 and No. 101053033). We thank Bastien Bedos for his help and commitment during his short visit in the team. 
\end{acknowledgements}

%
%
\bibliographystyle{aa} 
\bibliography{diazo} 

\begin{appendix} 
\section{Benchmark of Quantum Chemistry Calculations}

Here, we report the ability of quantum chemistry in correctly describing the different reaction products of N$_2$ + CH$_3^+$. In Table~\ref{table:reactenergy}, we report the energies obtained by the coupled-cluster (CC) theory and a DFT functional (B2PLYP-D3 which was used in recent works in astrochemistry with good performances \citep{bodo2019formation,skouteris2018genealogical}): they largely differ and, in particular, the DFT overestimates products stabilization. In fact, the DFT overestimates the entrance channel. Tests with other functionals reveal a similar behavior. \par

\begin{table*}
\caption{Reaction energies (in kcal/mol) for different products. In parenthesis we report the values corrected by ZPE. The zero is set on the reactants: CH$_3$ + N$_2^+$.}             
\label{table:reactenergy}      
\centering          
\begin{tabular}{l | c | c | c | c | c }      
\hline
Products         & CCSD(T)/aug-cc-pVQZ//    & B2PLYPD3/     & CCSD/     & CCSD(T)/  & ExFCI/ \\
         & CCSD(T)/cc-pVDZ$^{(a)}$   & aug-cc-pVTZ   & cc-pVDZ   & cc-pVDZ   & cc-pVDZ   \\
\hline   
$^1$N$_2$CH$_3^+$   & -180.21   & -207.50               & -174.90       & -173.25       & -171.47 \\
                    & (-173.37) & (-200.56)             & (-168.06)     &               &           \\
$^3$N$_2$CH$_3^+$   & -52.93    & -82.48                & -51.47        & -50.07        & -52.51    \\
                    & (-49.05)  & (-79.81)              & (-47.59)      &               &       \\
N$_2$ + CH$_3^+$    & -133.58   & -160.07               & -131.68       & -126.62       & -126.58 \\
                    & (-132.09) & (-158.42)             & (-130.19)     &               &   \\
$^2$NNCH$_2^+$ + $^2$H & -68.03 & -96.71                & -67.06        & -64.63        & -62.51 \\
                    & (-70.39)  & (-98.60)              & (-69.42)      &               &   \\
$^2$HNCNH$^+$ + $^2$H & -66.62  & -97.71                & -62.05        & -60.41        & -59.28 \\
                    & (-68.60)$^{(b)}$  & (-100.85)       & (-64.04)$^{(b)}$    &               & \\
$^2$Cyanamide$^+$ + $^2$H  & -64.18 & -94.08            & -62.43        & -59.94        & -56.72 \\
                    & (-65.44)  & (-94.98)              & (-63.70)      &               & \\
$^2$HCNNH$^+$ + $^2$H & -41.14  &   -71.04              & -37.19        & -33.92        & -32.53 \\
                    & (-44.09)  & (-73.57)              & (-40.14)      &               & \\
$^2$Diazirine + $^2$H & -24.70  & -52.06                & -21.55        & -19.17        & -17.54 \\
                    & (-27.25)  & (-54.69)              & (-24.10)      &               & \\
\hline
\end{tabular}
\\
$^{(a)}$ ZPE was calculated at CCSD/cc-pVDZ level of theory.
$^{(b)}$ MP2/aug-cc-pVTZ ZPE correction.
\end{table*}

\begin{table*}
\caption{Reaction energies (in kcal/mol) for different pathways. In parentheses, we report the ZPE-corrected values.}
\label{tab:react2}
\centering
\begin{tabular}{l|c | c | c} 
\hline
Reaction                                                & B2PLYPD3/aug-cc-pVTZ & CCSD/cc-pVDZ & CCSD(T)/cc-pVDZ \\
\hline
 N$_2$ + CH$_3^+$~$\rightarrow$~N$_2$CH$_2^+$ + H       & +63.36                & +64.61        & +61.99  \\
                                                        & (+59.82)              & (+60.76)      & \\
N$_2$ + CH$_3^+$~$\rightarrow$~N$_2$CH$_3^+$            & -47.44                & -43.22        & -46.63 \\
                                                        & (-42.14)              & (-37.87)      & \\
 N$_2$ + CH$_4^+$~$\rightarrow$~N$_2$CH$_2^+$ + H$_2$   & -1.36                 & -1.55         & -2.26 \\
                                                        & (-2.41)               & (-3.40)       & \\
 N$_2$ + CH$_4^+$~$\rightarrow$~N$_2$CH$_3^+$ + H       & -3.73                 & -5.78         & -7.27 \\
                                                        & (-2.31)               & (-4.69)       & \\
 N$_2$ + CH$_4^+$~$\rightarrow$~(N$_2$:H:CH$_3$)$^+$    & -16.68                & -16.56        & -17.37 \\
                                                        & (-14.88)              & (-14.78)      & \\
 N$_2$ + CH$_4^+$~$\rightarrow$~(HN$_2$:CH$_3$)$^+$     & -26.93                & -28.60        & -28.36 \\
                                                        & (-18.67)              & (-20.71)      & \\
 N$_2$ + CH$_4^+$~$\rightarrow$~TS                      & -0.08                 & +1.89$^{(a)}$ & +1.67$^{(a)}$ \\
                                                        & (+3.00)               & (+4.42)       & \\
\hline
\end{tabular}
\\
$^{(a)}$ on B2PLYPD3 geometry
\end{table*}

The CC theory is much more reliable, however in principle strongly correlated systems cannot be described correctly. At this end, we performed single point calculations using the ExFCI method. In Table~\ref{table:reactenergy} we report electronic energy values for the different reactions, which are very similar to what obtained from CCSD and CCSD(T) calculations performed with the same basis set (cc-pVDZ). Furthermore, the T2 values reported in Table~\ref{table:T2} are small, indicating that the strength of electron-electron correlation is weak and therefore that the CC theory gives reliable results.  Regarding the basis set dependence of CC calculation, it has been shown that an error of about 2 kcal/mol is expected with respect to complete basis set limit in the case of purely organic molecules \citep{loos2019density}, which is an accuracy affordable for the present study. \par 

\begin{table}
\caption{Dominant part of the ExFCI wave-function (T2 test). When T2~$<$~0.3-0.4 the correlation is considered to be weak and the coupled-cluster
method is reliable.}             
\label{table:T2}      
\centering          
\begin{tabular}{l | c }     
\hline
Species         & T2 on ExFCI wave-function \\
\hline                  
N$_2^+$         & 0.10 \\
CH$_3$          & 0.05 \\
N$_2$           & 0.09 \\
CH$_3^+$        & 0.06 \\
$^1$N$_2$CH$_3^+$   & 0.08 \\
$^3$N$_2$CH$_3^+$   & 0.10 \\
$^2$NNCH$_2^+$  & 0.10 \\
$^2$HNCNH$^+$   & 0.15 \\
$^2$Cyanamide$^+$ & 0.10 \\
$^2$HCNNH$^+$   & 0.09 \\
$^2$Diazirine$^+$ & 0.10 \\
\hline
\end{tabular}
\end{table}

The results clearly indicate that CC calculations are correct and can be taken as a solid basis. Unfortunately, as we can see from Table~\ref{table:reactenergy}, DFT provides different results, with large energy differences (between 20 and 30 kcal/mol) compared to accurate calculations. We have thus used CC for energies, frequencies and for all kinetic calculations.

\begin{figure*}
\includegraphics[width=8cm]{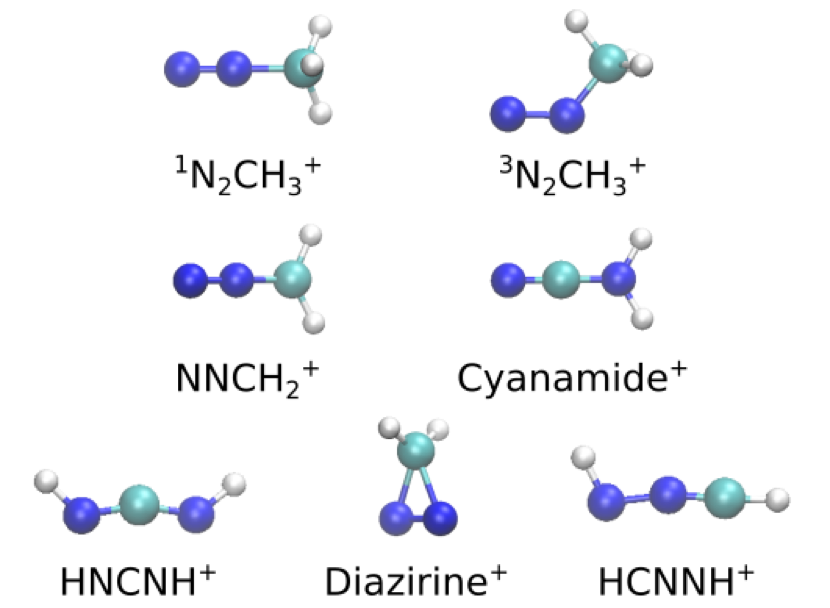}
\caption{Structures of the optimized geometries: N$_2$CH$_3^+$ and N$_2$CH$_2^+$ (for this last they are doublet spin states). Carbon atoms are in cyan, nitrogen in blue and hydrogen in white.}
\label{FigS1}
\end{figure*}

\begin{figure*}
\centering
\includegraphics[width=8cm]{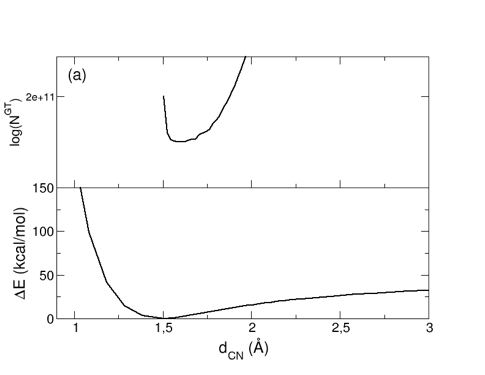}
\includegraphics[width=8cm]{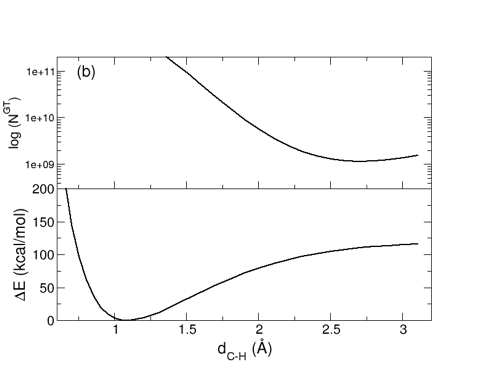}
\caption{Internal energy and sum of rovibrational states as a function of the reaction coordinate for the two reactions studied as obtained from CCSD/cc-pVDZ energies and projected frequencies. $N^{GT}$ values are reported for an internal energy of 174 kcal/mol.}
\label{FigS2}
\end{figure*}

\section{Long-range charge transfer}

We considered the possibility of long-range charge transfer from $N_2^+$ to CH$_3$ by 
evaluating the non-adiabiatic coupling matrix between two states at different charge localization.
To this end, we  considered the (N$_2$--CH$_3$)$^+$ system at fixed distance and calculated 
the excited states and their properties. Excited states calculations were performed using 
CIS formalism with cc-pVDZ basis set and 1s frozen core as implemented in Quantum Package~\citep{garniron2019quantum}. 
The 50 electronic excited states were calculated with both singlet and triplet spin multiplicity.
Then, the charge localization was done using Mulliken population analysis on the wave-function
of each state, as well as dipole moments of the states and transition dipole moments between
the states. 
The non-adiabatic couplings ($H_{ab}$) were thus obtained using the Generalized Mulliken-Hush method~\citep{CaveNewtonCPL,CaveNewtonJCP}: 

\begin{eqnarray}
|H_{ab}| &=& \frac{|\Vec{\mu}_{ij}| \Delta E_{ij}}{|\Delta \mu_{ab}|} ,\\
|\Delta \mu_{ab}| &=& \sqrt{(\mu_{ii}-\mu_{jj})^2 + 4 (\mu_{ij})^2 }
,\end{eqnarray}

where $a$ and $b$ indicate the states with localized charge (diabatic states), while $i$ and $j$ show the adiabatic states, $\mu_{ij}$ are the transition dipole moments between two adiabatic states,
$\Delta E_{ij}$ is the energy difference and $\mu_{ii}$ is the dipole moment of the state, $i$. This approach provides a simple diabatization with results similar to more elaborated ones~\citep{CaveNewtonJCP} and
it is very efficient when there is a clear and different charge localization.

To have an evaluation of the possibility of long-range charge transfer we considered a configuration in which the N--C distance is 5.5~\mbox{\AA,} 
which is surely longer than those of the variational transition state(s).
As expected, the ground state is a singlet with the following
charge distribution: N$_2$ + CH$_3^+$. 
The first excited states with inverted charge localization (N$_2^+$ + CH$_3$) are at a level of 8.78~eV in energy and are four degenerate singlet and triplet states with zero or very small non-adiabatic coupling with the ground state. Details on the lowest lying excited states with the charge on the N$_2$ group are summarized in Table~\ref{table:NonAd}, together with the non-adiabatic couplings with the ground state. 
At it can be seen, the values of H$_{ab}$ are zero or very small, thus, we can consider that at a first approximation, the long-range charge transfer is negligible.

\begin{table}
\caption{Properties of the low-lying excited states with charge on the N$_2$ group for the (N$_2$--CH$_3$)$^+$ system at a 5.48~\mbox{\AA} N--C distance.}             
\label{table:NonAd}      
\centering          
\begin{tabular}{l | ccc }     
\hline
State         & Energy (eV) & Multiplicity & H$_{ab}$ (mH) \\
\hline                  
13           & 8.78     & |T$\rangle$   & 0 \\
14           & 8.78     &  |T$\rangle$ & 0 \\
15           & 8.78     & |S$\rangle$ & 0.009 \\
16          &  8.78     & |S$\rangle$ & 0.009 \\      
19          & 9.38      & |T$\rangle$   & 0 \\
20          & 9.38      & |S$\rangle$   & 0.4 \\
\hline
\end{tabular}
\end{table}

\end{appendix}

\end{document}